\documentclass[11pt]{article}
\usepackage{latexsym}
\usepackage{mathptmx}
\usepackage[T1]{fontenc}

\usepackage[margin=1in]{geometry}  % set the margins to 1in on all sides

\usepackage{amsmath}
\usepackage{amsfonts}
\usepackage{amssymb}
\usepackage{amsbsy}
\usepackage{amsthm}
\usepackage{bm}
\usepackage{multirow, booktabs}
\usepackage{algorithm}
\usepackage{algorithmic}
\usepackage{enumitem}
\usepackage{epstopdf}
\usepackage{graphicx}
\usepackage{subcaption}
\usepackage{booktabs}
\usepackage{xcolor}
\usepackage{url}
\usepackage[colorlinks=true, linkcolor=blue, citecolor=blue]{hyperref}

\theoremstyle{remark}

\newcommand{\cov}{\mbox{cov}}

\newcommand{\var}{\mbox{var}}

\DeclareMathOperator*{\E}{\mathbb{E}}
\title{Adaptive Nystr\"om for Gaussian Process Regression}

\author{Lulu Kang\\
Department of Mathematics and Statistics\\
University of Massachusetts Amherst\\
lulukang@umass.edu
}

\date{}
\begin{document}

\maketitle

\vspace{-12pt}

\section*{ABSTRACT}
Gaussian Process Regression (GPR) is a robust framework for uncertainty quantification, yet its $O(n^3)$ complexity limits its scalability. 
Low-rank Nystr\"om approximations can reduce this burden to $O(nm^2)$, but their accuracy depends heavily on the selection of landmark points. 
We propose an adaptive Nystr\"om approach that greedily selects landmarks to minimize the trace residual of the kernel approximation error. 
Unlike static approximations, our method interleaves landmark expansion with hyperparameter optimization, allowing the selection process to adapt as the covariance structure is refined. 
Numerical experiments on five benchmark functions demonstrate that this method significantly outperforms random landmark selection in both accuracy and stability. 
It achieves predictive performance comparable to exact GP inference while maintaining linear scaling with respect to the sample size, providing a principled and efficient framework for large-scale computer experiments. 

\section{INTRODUCTION}
\label{sec:intro}

Gaussian Process Regression (GPR) \cite{williams2006gaussian,santner2003design} is a powerful non-parametric framework widely used in machine learning, statistics, and scientific computing for regression and uncertainty quantification tasks. 
Its flexibility and ability to provide probabilistic predictions make it a popular choice in applications ranging from geostatistics \cite{cressie2015statistics} and climate modeling \cite{donnelly2024forecasting} to robotics \cite{6654139} and healthcare \cite{Chung_Kim_Lee_Kim_Hwang_Yang_2020}. 
However, the computational complexity of GPR, $O(n^3)$, which scales cubically with the number of data points due to the need for matrix inversion and Cholesky decomposition, poses a significant challenge for large-scale datasets. 
This limitation has spurred extensive research into scalable approximations, such as sparse GPR \cite{pmlr-v37-sheth15,pmlr-v2-snelson07a}, low-rank approximations \cite{Katzfuss2017,Song13112024}, and distributed computing \cite{pmlr-v37-deisenroth15}. 
Despite these advances, the demand for faster and more memory-efficient GPR implementations continues to grow, especially in high-dimensional and large-scale settings. 

Among the various low-rank techniques, the Nystr\"{o}m method \cite{williams2000using,drineas2005nystrom} has emerged as a cornerstone for scaling kernel-based models by approximating the full kernel matrix through a subset of its columns, often referred to as landmark points. 
While the computational efficiency of the Nystr\"om method is well-documented, its accuracy is inherently sensitive to the choice of these landmarks. 
In practice, uniform random selection is the most common strategy due to its minimal overhead \cite{fowlkes2004spectral,NIPS2002_5d6646aa}; however, such a stochastic approach often fails to capture the intrinsic structural nonuniformity of the kernel matrix, leading to inconsistent approximations. 
To address this limitation, \cite{645529.657980} proposed a greedy column selection algorithm providing a low-rank kernel matrix approximation, \cite{bach2002kernel} introduced the Incomplete Cholesky decomposition (ICL) sampling, and \cite{NIPS2009_a49e9411,kumar2012sampling} proposed a family of ensemble-based sampling method based on reconstruction error. 
Many other fixed and adaptive sampling methods have been created for the Nystr\"om method. 
See \cite{sun2015review} for a comprehensive review.

In this paper, we propose an adaptive Nystr\"{o}m approach for Gaussian process regression. 
Our method utilizes a greedy, sequential selection strategy based on the trace residual of the kernel approximation error. 
By interleaving the expansion of the landmark set with hyperparameter optimization, the selection process adaptively identifies the most informative design points as the covariance structure is refined. This framework provides a principled, linear-scaling alternative to exact GP inference, significantly reducing the computational burden while maintaining high predictive fidelity.

The remainder of this paper is organized as follows. Section 2 provides a preliminary review of Gaussian process regression and the associated computational challenges involving the marginal likelihood. 
Section 3 details our proposed methodology, including the Woodbury-based reductions for the Nystr\"om likelihood, the trace-residual selection criterion, and the integrated sequential estimation algorithm. 
In Section 4, we evaluate the performance of the adaptive Nystr\"{o}m method through a series of numerical experiments using several benchmark test functions, focusing on convergence properties and computational efficiency. 
Finally, Section 5 provides concluding remarks and directions for future research.

\section{Preliminary: Gaussian Process Regression}

We first briefly review GPR in its popular form. 
Denote $\{\bm x_i, y_i\}_{i=1}^n$ as the $n$ pairs of input and output data from a certain computer experiment, and $\bm x_i\in \Omega \subseteq \mathbb{R}^d$ are the $i$th experimental input values and $y_i\in \mathbb{R}$ is the corresponding output. 
In this paper, we only consider the case of univariate response. 
GPR is built on the following model assumption of the response, 
\begin{equation}\label{eq:model}
y_i=\mu(\bm x)+ Z(\bm x_i)+\epsilon_i, \quad i=1,\ldots, n,
\end{equation}
where $\mu$ is either set to be $0$ (when $\bm y$ is centered in pre-processing), an unknown constant, or a linear combination of pre-specified basis functions, $\mu(\bm x)=\bm g(\bm x)^\top \bm \beta$, which is what we follow here. 
The random noise $\epsilon_i$'s are independently and identically distributed following $\mathcal{N}(0, \sigma^2)$. 
They are also independent of the other stochastic components of \eqref{eq:model}. 
We assume the GP prior on the stochastic function $Z(\bm x)$, which is denoted as $Z(\cdot)\sim GP(0, \tau^2 K)$, i.e., $\mathbb{E}[Z(\bm x)]=0$ and the covariance function
\[
\cov[Z(\bm x_1), Z(\bm x_2)]=\tau^2 K(\bm x_1, \bm x_2;\bm \theta).
\]
In most applications, we use the stationary assumption of $Z(\bm x)$, and thus the variance $\tau^2$ is a constant. 
The function $K(\cdot, \cdot;\bm \theta): \Omega \times \Omega \mapsto \mathbb{R}_{+}$ is the correlation of the stochastic process with hyperparameters $\bm \theta$. 
For it to be valid, $K(\cdot,\cdot;\bm \theta)$ must be a symmetric positive definite kernel function. 
Gaussian and Mat\'{e}rn kernel functions are among the most popular used ones and their definitions are
\begin{align*}
\text{Gaussian }& K(\bm x_1, \bm x_2; \bm \theta)=\exp\left\{-\sum_{j=1}^d \theta_j(x_{1j}-x_{2j})^2\right\}, \\
\text{Mat\'{e}rn }& K(\bm x_1, \bm x_2; \bm \theta, \nu)\propto \prod_{j=1}^d \left(\theta_j|x_{1j}-x_{2j}|\right)^{\nu} B_{\nu}(\sqrt{2\nu\theta_j}|x_{1j}-x_{2j}|),
\end{align*}
with $\bm \theta \in \mathbb{R}^d$ and $\bm \theta\geq 0$. 
The two kernels are anisotropic in the sense that $\theta_j$'s are different for different dimensions. 
In isotropic kernels, or RBF kernels, $\theta_j$'s are the same for all dimensions. 

In terms of response $Y(\bm x)$, it follows a Gaussian process with the following mean and covariance, 
\begin{align*}
\mathbb{E}[Y(\bm x)]&=\mu(\bm x), \quad \forall \bm x \in \Omega\\
\cov[Y(\bm x_1), Y(\bm x_2)]&=\tau^2 K(\bm x_1, \bm x_2;\bm \theta) + \sigma^2\delta(\bm x_1, \bm x_2), \quad \forall \bm x_1, \bm x_2\in \Omega,\\\
&=\tau^2\left[ K(\bm x_1, \bm x_2;\bm \theta) + \eta\delta(\bm x_1, \bm x_2)\right], 
\end{align*}
where $\delta(\bm x_1,\bm x_2)=1$ if $\bm x_1=\bm x_2$ and 0 otherwise, and $\eta=\sigma^2/\tau^2$. 
So $\eta$ is interpreted as the noise-to-signal ratio. 
For deterministic computer experiments, the noise component is not part of the model, i.e., $\sigma^2=0$ and $\eta=0$. 
However, a nugget effect, which is a small $\eta$ value, is usually included in the covariance function to avoid the singularity of the covariance matrix \cite{peng2014on}. 
The unknown parameter values of the GP model are $(\bm \beta, \bm \theta, \tau^2, \eta)$. 

Following the empirical Bayes or maximum likelihood estimation, $\mu$ and $\tau^2$ have the tractable solutions
\begin{equation*}
\hat{\bm \beta} =\left[\bm G^\top (\bm K+\eta\bm I_n)^{-1}\bm G\right]^{-1}\left[\bm G^\top (\bm K+\eta\bm I_n)^{-1}\right]\bm y, \quad 
\hat{\tau}^2 =\frac{1}{n}(\bm y-\bm G\hat{\bm \beta})^\top (\bm K+\eta \bm I_n)^{-1} (\bm y-\bm G\hat{\bm \beta}),
\end{equation*}
where $\bm y$ is the vector of output observations, $\bm I_n$ is an identity matrix of size $n$, $\bm G$ is a $n\times p$ matrix with $\bm g(\bm x_i)^\top$ as the $i$-th row, $\bm K$ is the $n\times n$ kernel matrix whose entries are $\bm K_{ij}=K(\bm x_i, \bm x_j)$. 
Therefore, $\bm K$ is a symmetric positive definite matrix since $K$ is such a kernel. 
To estimate the remaining parameters $\eta$ and $\bm \theta$, we replace $\bm \beta$ and $\tau^2$ by estimates in the likelihood, and maximize the updated likelihood, which is equivalent to solving
\[\min_{\eta,\bm \theta} n\log(\hat{\tau}^2)+\log\det(\bm K+\eta\bm I_n).\]
The prediction at any query point $\bm x$ conditioned on the observed data are the conditional mean
\[\hat{y}(\bm x)=\E(y(\bm x)|\bm y)=\bm g(\bm x)^\top\hat{\bm \beta}+\bm k(\bm x)(\bm K+\hat{\eta} \bm I_n)^{-1}(\bm y-\bm G\hat{\bm \beta}),\]
and the conditional variane is 
\[
\var(y(\bm x)|\bm y)=\tau^2\left\{1-\bm k(\bm x)^\top (\bm K+\eta \bm I_n)^{-1}\bm k(\bm x)+\bm c(\bm x)^\top \left[\bm G^\top(\bm K+\eta\bm I_n)^{-1}\bm G\right]^{-1}\bm c(\bm x)\right\},
\]
where $\bm c(\bm x)=g(\bm x)-\bm G^\top (\bm K+\eta\bm I_n)^{-1}\bm k(\bm x)$. 

\section{The Adaptive Nystr\"om Method}
\label{sec:nystrom}

The Nystr\"om method is a widely used technique for constructing a low-rank approximation of a kernel matrix, reducing the computational burden of GPR from $O(n^3)$ to $O(nm^2)$. 
It relies on a set of $m \ll n$ landmark points (or inducing inputs) $\mathcal{Z} = \{\bm{z}_1, \dots, \bm{z}_m\} \subset \mathcal{X}$, where $\mathcal{X}$ is the set of $n$ design points. 
Define the cross-kernel matrix $\bm{C} = \bm{K}_{XZ} \in \mathbb{R}^{n \times m}$ with entries $C_{ij} = K(\bm{x}_i, \bm{z}_j; \bm{\theta})$, and the landmark kernel matrix $\bm{W} = \bm{K}_{ZZ} \in \mathbb{R}^{m \times m}$ with entries $W_{ij} = K(\bm{z}_i, \bm{z}_j; \bm{\theta})$. 
Assuming $\bm{W}$ is invertible, the rank-$m$ Nystr\"om approximation of the full kernel matrix is
\[
\tilde{\bm{K}} = \bm{C} \bm{W}^{-1} \bm{C}^\top.
\]
Geometrically, this approximation corresponds to an orthogonal projection of the canonical feature maps onto the subspace spanned by the landmark features \cite{williams2000using,zhang2008improved,wild2023connections}. 
While the most common technique for landmark selection is uniform random sampling due to its minimal overhead \cite{sun2015review,li2010making}, such a strategy often ignores the structural nonuniformity of the kernel matrix, potentially leading to poor approximations unless $m$ is large.

\subsection{Nystr\"om-based Likelihood and Estimation}
\label{sec:nystrom_likelihood}

By replacing the full kernel matrix $\bm{K}$ with its Nystr\"om approximation $\tilde{\bm{K}}$, we define the approximate covariance matrix $\tilde{\bm{R}} = \tilde{\bm{K}} + \eta \bm{I}_n$. This structure allows us to avoid the inversion of the $n \times n$ matrix by applying the Woodbury matrix identity \cite{woodbury1950inverting}:
\[
\tilde{\bm{R}}^{-1} = \frac{1}{\eta} \bm{I}_n - \frac{1}{\eta^2} \bm{C} \bm{M}^{-1} \bm{C}^\top,
\]
where $\bm{M} = \bm{W} + \eta^{-1} \bm{C}^\top \bm{C} \in \mathbb{R}^{m \times m}$. The log-determinant is similarly reduced via the matrix determinant lemma \cite{harville2008matrix} to
\[
\log \det \tilde{\bm{R}} = n \log \eta + \log \det \bm{M} - \log \det \bm{W}.
\]
These reductions ensure that the negative log marginal likelihood $\tilde{\mathcal{L}}_m(\bm{\theta}, \eta)$ can be evaluated in $O(nm^2 + m^3)$ time. To further accelerate the estimation, we profile out $\tau^2$ and $\bm{\beta}$. In the constant mean case ($\mu(\bm{x}) = \beta$), the profiled estimates are
\[
\hat{\beta} = \frac{\bm{1}_n^\top \tilde{\bm{R}}^{-1} \bm{y}}{\bm{1}_n^\top \tilde{\bm{R}}^{-1} \bm{1}_n}, \quad \hat{\tau}^2 = \frac{1}{n} \Bigg( \frac{\|\bm{y} - \hat{\beta}\bm{1}_n\|^2}{\eta} - \frac{1}{\eta^2} \bm{c}^\top \bm{M}^{-1} \bm{c} \Bigg),
\]
where $\bm{c} = \bm{C}^\top (\bm{y} - \hat{\beta}\bm{1}_n)$. Similar Woodbury-based reductions apply to the polynomial mean case ($\mu(\bm{x}) = \bm{g}(\bm{x})^\top \bm{\beta}$), where the $q \times q$ system for $\hat{\bm{\beta}}$ is solved in $O(q^3)$ after computing $\bm{G}^\top \tilde{\bm{R}}^{-1} \bm{G}$ and $\bm{G}^\top \tilde{\bm{R}}^{-1} \bm{y}$ in $O(nmq + m^2q)$ time. The resulting approximate objective is
\begin{equation}\label{eq:nys_objective}
\tilde{\mathcal{L}}_m(\bm{\theta}, \eta) = n \log \hat{\tau}^2 + n \log \eta + \log \det \bm{M} - \log \det \bm{W}.
\end{equation}

\subsection{Adaptive Landmark Selection via Trace Residual}
\label{sec:landmark_selection}

The accuracy of $\tilde{\mathcal{L}}_m$ as a surrogate for the exact marginal likelihood depends critically on the landmark set $\mathcal{Z}$. Rather than relying on random selection, we propose a greedy, sequential strategy that adds landmarks to minimize the approximation error. For a set of landmarks $\mathcal{Z}$, the residual variance at a point $\bm{x}_i$ is
\[
r(\bm{x}_i) = K(\bm{x}_i, \bm{x}_i) - \bm{k}_Z(\bm{x}_i)^\top \bm{W}^{-1} \bm{k}_Z(\bm{x}_i),
\]
where $\bm{k}_Z(\bm{x}_i) = (K(\bm{x}_i, \bm{z}_1), \dots, K(\bm{x}_i, \bm{z}_m))^\top$. This quantity represents the squared norm of the residual after projecting the feature $\phi(\bm{x}_i)$ onto the subspace spanned by the landmarks \cite{wild2023connections}. 
The total kernel approximation error is given by the trace $\operatorname{tr}(\bm{K} - \tilde{\bm{K}}) = \sum_{i=1}^n r(\bm{x}_i)$. 

The choice of the trace residual as a selection criterion is principled; the Kullback--Leibler divergence between the exact GP posterior and the sparse variational posterior is bounded above by a function of $\operatorname{tr}(\bm{K} - \tilde{\bm{K}})$ \cite{wild2023connections,titsias2009variational}. By greedily selecting the point $\bm{x}^*$ that maximizes $r(\bm{x}_i)$, we maximally decrease this trace at each step:
\begin{equation}\label{eq:greedy}
\bm{x}^* = \operatorname*{arg\,max}_{\bm{x}_i \notin Z} \; \left( 1 - \big\| \bm{L}_W^{-1} \bm{k}_Z(\bm{x}_i) \big\|^2 \right),
\end{equation}
where $\bm{L}_W$ is the Cholesky factor of $\bm{W}$. This selection ensures that the landmark set ``covers'' the regions of highest uncertainty in the low-rank approximation, leading to more robust hyperparameter estimates than random selection.

\subsection{Sequential Estimation and Prediction}

The complete estimation procedure interleaves greedy landmark expansion with hyperparameter optimization, as detailed in Algorithm~\ref{alg:nystrom_gpr}. We begin with a small random subset $m_0$ and iteratively add landmarks until the relative change in $\bm{\theta}$ ($\Delta_\theta^{(t)}$) and the mean trace residual ($\bar{r}^{(t)}$) fall below user-specified tolerances $\delta_\theta$ and $\delta_r$.

Once the model is fitted, predictions for a test point $\bm{x}_*$ are computed without forming $n \times n$ matrices. The predictive mean and variance are
\[
\hat{y}(\bm{x}_*) = \bm{g}(\bm{x}_*)^\top \hat{\bm{\beta}} + \bm{k}_Z(\bm{x}_*)^\top \bm{\beta}^*, \quad \text{where } \bm{\beta}^* = \frac{1}{\hat{\eta}} \bm{M}^{-1} \bm{C}^\top (\bm{y} - \bm{G}\hat{\bm{\beta}}),
\]
\[
\operatorname{var}(y(\bm{x}_*) \mid \bm{y}) \approx \hat{\tau}^2 \Big( 1 + \hat{\eta} - \bm{k}_Z(\bm{x}_*)^\top \bm{W}^{-1} \bm{k}_Z(\bm{x}_*) + \frac{1}{\hat{\eta}} \bm{k}_Z(\bm{x}_*)^\top \bm{M}^{-1} \bm{k}_Z(\bm{x}_*) \Big).
\]
The total complexity of the sequential procedure is $O(T \cdot E \cdot n m^2)$, where $T$ is the number of landmark updates and $E$ is the number of likelihood evaluations per optimization. For $n \gg m$, this linear scaling in $n$ allows GPR to be applied to datasets where exact $O(n^3)$ inference is computationally infeasible.
The tuning parameters $m_0$ and $m_{\max}$ should be selected based on the available computational time and resources. 
These tolerances $\delta_{\theta}$ and $\delta_r$ directly govern the trade-off between computational runtime and predictive accuracy. 
Setting $\delta_r$ too high (e.g., $\geq 10^{-2}$) terminates the expansion prematurely, resulting in a sparse landmark set that may not fully capture local features. 
Conversely, setting $\delta_r$ too low (e.g., $\leq 10^{-6}$) leads to over-allocation of landmarks, pushing the algorithm toward $m_{\max}$ and increasing the computational cost with minimal gains in accuracy. 
The hyperparameter tolerance $\delta_{\theta}$ acts as a secondary safeguard; values between $10^{-3}$ and $10^{-4}$ generally ensure that the covariance structure has stabilized before stopping. 
Across our experiments, the defaults of $\delta_{\theta}=10^{-3}$ and $\delta_r=10^{-4}$ consistently provided a stable balance, achieving near-exact GP accuracy while avoiding unnecessary sequential optimization loops.

\begin{algorithm}[tb]
\caption{Adaptive Nystr\"om Gaussian Process Regression}
\label{alg:nystrom_gpr}
\begin{algorithmic}[1]
\REQUIRE Training data $(\bm{X}, \bm{y})$, initial landmark count $m_0$, max landmarks $m_{\max}$, tolerances $\delta_\theta$, $\delta_r$.
\ENSURE Estimated parameters $(\hat{\tau}^2, \hat{\bm{\beta}})$, hyperparameters $(\hat{\bm{\theta}}, \hat{\eta})$, and landmark set $Z$.
\STATE Initialize $Z$ by sampling $m_0$ landmarks uniformly at random from $\bm{X}$.
\STATE Obtain initial $\hat{\bm{\theta}}^{(0)}, \hat{\eta}^{(0)}$ by minimizing $\tilde{\mathcal{L}}_{m_0}(\bm{\theta}, \eta)$ using BOBYQA.
\FOR{$t = 1$ \TO $m_{\max} - m_0$}
    \STATE Compute residual variances $\{r(\bm{x}_i)\}_{i=1}^n$ using the current $\hat{\bm{\theta}}^{(t-1)}$.
    \STATE $\bm{x}^* \leftarrow \arg\max_{\bm{x}_i \notin Z} r(\bm{x}_i)$; update landmark set $Z \leftarrow Z \cup \{\bm{x}^*\}$.
    \STATE Update $\hat{\bm{\theta}}^{(t)}, \hat{\eta}^{(t)} \leftarrow \arg\min \tilde{\mathcal{L}}_{m_t}(\bm{\theta}, \eta)$ (warm-start from previous estimates).
    \STATE Compute convergence metrics $\Delta_\theta^{(t)}$ and $\bar{r}^{(t)}$.
    \IF{$\Delta_\theta^{(t)} < \delta_\theta$ \AND $\bar{r}^{(t)} < \delta_r$}
        \STATE \textbf{break}
    \ENDIF
\ENDFOR
\STATE Compute final variance $\hat{\tau}^2$ and coefficients $\hat{\bm{\beta}}$.
\RETURN $\hat{\bm{\theta}}, \hat{\eta}, \hat{\tau}^2, \hat{\bm{\beta}}, Z$.
\end{algorithmic}
\end{algorithm}

\subsection{Computational Complexity}
\label{sec:complexity}

Each evaluation of the Nystr\"om likelihood~\eqref{eq:nys_objective} requires $O(n m^2 + m^3)$ operations rather than $O(n^3)$. The sequential algorithm performs $T$ such optimizations, where $T$ is typically small (fewer than 20 in our experiments). The residual variance computation costs $O(n m^2)$ per iteration, and the landmark update costs $O(m^3)$ if the Cholesky factor is recomputed from scratch (or $O(m^2)$ if updated via rank-one modifications).

The dominant cost remains $O(T \cdot E \cdot n m^2)$, where $E$ is the number of likelihood evaluations per optimization (typically 100--300 with derivative-free methods). For $n \gg m$, this represents substantial savings over the $O(n^3)$ per evaluation required by exact GP inference, and the linear scaling in $n$ allows the method to scale to datasets where exact inference is infeasible. 

\subsection{Relationship to Existing Low-Rank and Sparse GP Methods}
\label{sec:relationship}

To clarify the positioning of the proposed adaptive Nystr\"{o}m method, we discuss its connections to several established paradigms in kernel approximation and sparse Gaussian processes. 

Compared with the pivoted cholesky factorization method \cite{higham2002accuracy}, the greedy selection criterion in \eqref{eq:greedy}, which maximizes the residual $r(\bm x_i)$, is mathematically equivalent to performing a pivoted Cholesky decomposition on the kernel matrix $\bm K$. 
While pivoted Cholesky is typically executed as a static algebraic step with pre-specified hyperparameters, our method differs by updating the kernel hyperparameters $\bm \theta$ during the selection process.

In GP active learning, greedy variance reduction seeks to add design points where the predictive uncertainty is highest. 
Because the trace residual $r(\bm x_i)=K(\bm x_i, \bm x_i)-\bm k_Z(\bm x_i)^\top \bm W^{-1}\bm k_Z(\bm x_i)$ is exactly the variance of the GP prior at $\bm x_i$ conditioned on the current landmark set $Z$, our greedy selection scheme corresponds to finding the point of maximum prior uncertainty under the current model.

Randomized Nystr\"{o}m techniques often sample columns based on ridge leverage scores \cite{sun2015review}, which reflect the relative importance of a column in representing the range space of the kernel. 
Unlike randomized leverage-score sampling, which is designed to provide global error bounds under random draws, our method uses a deterministic greedy strategy that actively targets local approximation errors.

In sparse GP models (e.g., \cite{titsias2009variational}), inducing inputs are often treated as continuous variational parameters optimized via gradient descent. Our method restricts the inducing points (landmarks) to be a subset of the observed training data $\mathcal{X}$. 
By avoiding continuous optimization over the $d\times m$ coordinate space, we simplify the optimization landscape while leveraging the discrete trace-residual reduction.

The primary novelty of our method is the interleaving of landmark selection and hyperparameter optimization (Algorithm \ref{alg:nystrom_gpr}). In standard low-rank approximations, hyperparameters are either fixed a priori or optimized after the landmark set has been determined. 
By coupling the two processes, the landmark selection dynamically adjusts as the lengthscales and noise-to-signal ratio are updated, ensuring that the landmarks are placed in areas of highest relevance under the most accurate covariance structure.

\section{Numerical Examples}

In this section, we evaluate the performance of the proposed adaptive Nystr\"om GPR method through several simulation experiments using benchmark test functions. 
The test functions are included in the Virtual Library of Simulation Experiments \cite{simulationlib}. 
All methodologies, including the conventional GPR and the Nystr\"om-accelerated variants, were implemented in the R programming language. 
While high-performance R packages such as \texttt{DiceKriging} \cite{DiceKriging} are widely available for Gaussian process modeling, we conducted our experiments using a unified custom framework. 
This approach ensures that the conventional GPR and the proposed adaptive Nystr\"{o}m method are compared on common ground, utilizing identical covariance kernel logic, similar log-likelihood evaluation routines, and optimization procedures. 
Consequently, any observed differences in accuracy or computational efficiency can be strictly attributed to the landmark selection and approximation strategies rather than differences in implementation or library-specific optimizations.

Experimental designs for all test cases were generated using the \texttt{lhs} package \cite{lhs}. 
For hyperparameter estimation, we utilized the \texttt{nloptr} package \cite{nloptr}, specifically employing the BOBYQA (Bounded Optimization BY Quadratic Approximation) \cite{powell2009bobyqa} algorithm. 
This derivative-free optimizer was configured with a maximum of 2,000 evaluations and a relative objective tolerance of $10^{-6}$ to ensure robust convergence across different datasets. 
In the adaptive selection process, the sequential refinement was governed by two convergence tolerances: the relative change in hyperparameter estimates was set to $\delta_{\theta} = 10^{-3}$, and the mean trace residual tolerance was set to $\delta_r = 10^{-4}$. 
These settings balance the fidelity of the low-rank approximation with the need for computational parsimony.

\subsection{Landmark Set Size} 

To evaluate the performance of the proposed adaptive Nystr\"{o}m GPR, we conduct an experiment using the Borehole test function \cite{simulationlib}, a well-known benchmark in computer experiments. 
The function models the flow of water through a borehole and is defined by eight input variables:
$$y = \frac{2\pi T_u (H_u - H_l)}{\ln(r/r_w) \left[ 1 + \frac{2 L T_u}{\ln(r/r_w) r_w^2 K_w} + \frac{T_u}{T_l} \right]}$$
The inputs represent physical properties such as transmissivity and hydraulic head, with dimensions $d=8$. 
This function is characterized by a generally smooth surface with strong localized dependencies, making it an ideal candidate for testing landmark selection strategies.

We generated a training dataset of $n = 1000$ points using a maximin Latin Hypercube Design (LHD). 
All eight input dimensions were scaled to the range $[-1, 1]$ to ensure numerical stability during hyperparameter optimization. 
A separate test set of $n_{test} = 1000$ points was generated via a different LHD to evaluate the prediction accuracy.

We compared three models:
\begin{enumerate}
\item Full GP Baseline: Conventional GPR using all $n=1000$ points (the theoretical gold standard).
\item Adaptive Nystr\"om (the proposed): Starting with $m_0 = 20$ initial random landmarks and sequentially adding points that maximize the trace residual $r(\mathbf{x}_i)$ as defined in Eq. (17).
The maximum landmark points is $m_{\max}=100$. 
\item Random Nystr\"om: landmarks selected uniformly at random from the training set, with hyperparameter optimization performed once for each $m$. 
This single-stage optimization ensures a strong, direct baseline for comparing predictive accuracy.
\end{enumerate} 
For all models, we employed a Mat\'{e}rn kernel with $\nu = 1.5$. 
Optimization was conducted via the BOBYQA algorithm. 
We examined two mean structures: Ordinary Kriging (constant mean) and Polynomial-Mean Kriging (polynomial mean including linear and quadratic terms). 
Accuracy was measured using the Standardized Root Mean Squared Prediction Error (S-RMSPE):
$$\text{S-RMSPE} = \frac{\sqrt{\frac{1}{n_{test}} \sum_{i=1}^{n_{test}} (y_i - \hat{y}_i)^2}}{\text{sd}(y_{test})}$$

\begin{figure}
\centering
\includegraphics[width=0.8\textwidth]{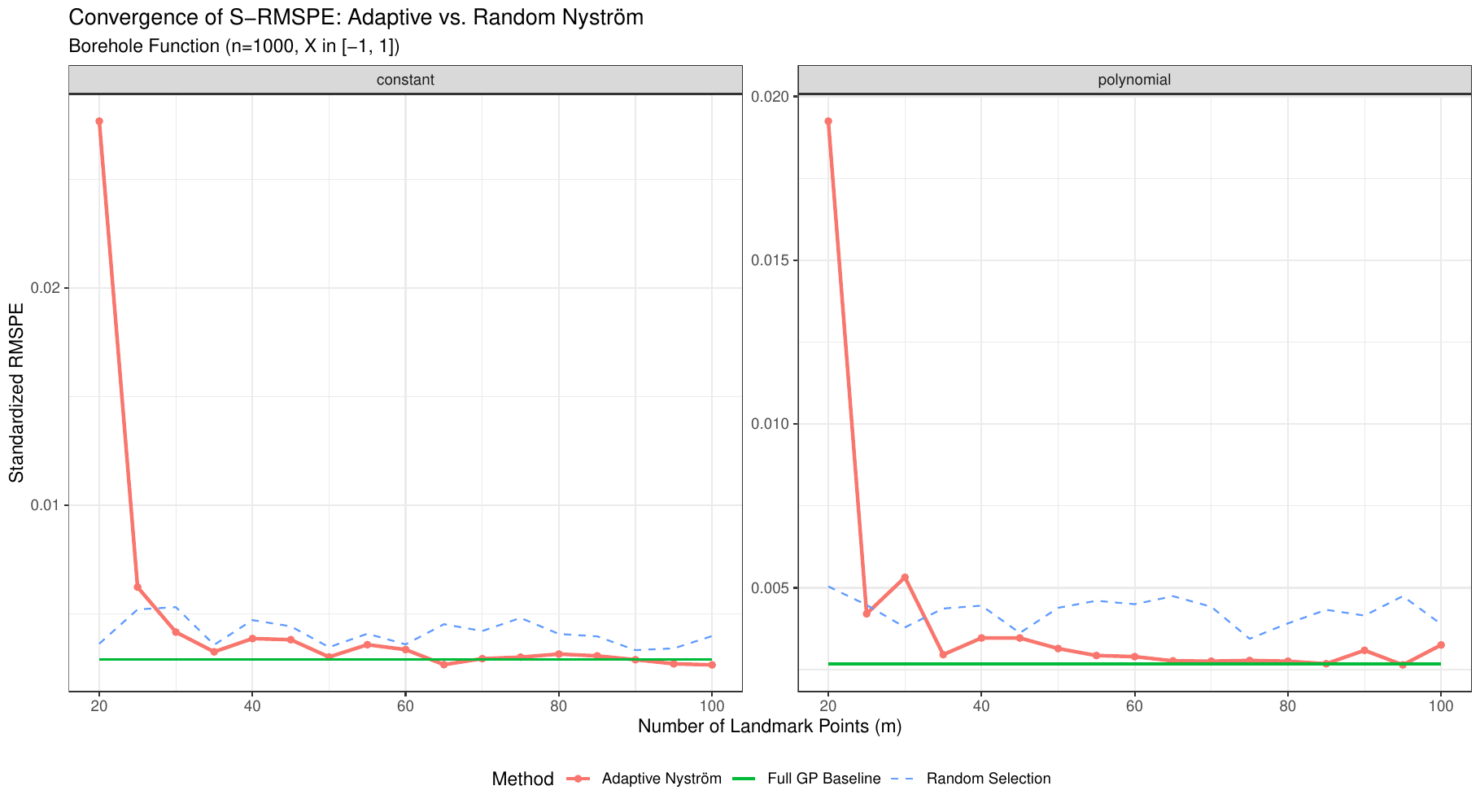}
\caption{Borehole Example: S-RMSPE of the three methods with respect to increasing number of landmark points.}
\label{fig:borehole}
\end{figure}

Figure \ref{fig:borehole} displays the convergence of S-RMSPE as the number of landmark points $m$ increases from 20 to 100. 
We make several observations based on the following aspects.
(1) Superiority of Adaptive Selection: In both the constant and polynomial mean cases, the adaptive Nystr\"{o}m approach (solid red line) demonstrates a much sharper initial decrease in error compared to random selection (dashed blue line). For instance, in the constant mean facet, the adaptive method effectively matches the performance of the Full GP baseline with as few as $m=35$ landmark points. 
(2) Stability vs. Volatility: The random selection approach exhibits significant volatility. Because it does not account for the geometric distribution or the residual variance of the points, it often fails to capture the essential features of the kernel space, leading to inconsistent prediction errors even as $m$ increases. In contrast, the adaptive method monotonically converges toward the Full GP baseline, reflecting the effectiveness of the greedy trace-residual criterion.
(3) Impact of Mean Structure: The polynomial mean models (right facet) achieve a lower overall S-RMSPE than the constant mean models. This indicates that the Borehole function possesses strong global trends that are well-captured by quadratic basis functions. Notably, the adaptive Nystr\"{o}m method excels in this setting as well; the ``gap'' between adaptive and random selection is even more pronounced here, as the adaptive method quickly identifies the sparse landmark set required to correct the global trend with local GP deviations.
(4) Computational Efficiency: While the Full GP (solid green line) represents the most accurate model, it requires $O(n^3)$ operations. The results show that the adaptive Nystr\"{o}m method achieves an identical level of accuracy using only $m=40$ points—just 4\% of the total training data. 
This results in a massive reduction in the computational footprint for both estimation and prediction, without a statistically significant loss in accuracy.

\subsection{Accuracy and Computing Time}

In this subsection, we provide a comprehensive comparison between the proposed adaptive Nystr\"om GPR, the random Nystr\"om baseline, and the full GPR model. 
The experiments were designed to evaluate the trade-off between predictive fidelity and computational cost across varying dimensions and sample sizes. 

For each test function, we fixed a training set of size $n$ and a large independent test set ($n_{test} = 1000$) generated via maximin LHD. 
To account for the stochastic nature of landmark initialization and optimization, we conducted 10 replications for both the adaptive and random Nystr\"om methods. 
In each replication, the methods were initialized with a different random set of $m_0$ landmarks. 
For the random Nystr\"om baseline, we used a fixed landmark set size equal to the number of landmarks used by the adaptive method upon convergence in the same replication to ensure a fair comparison of the selection strategy. 
Given the $O(n^3)$ cost of conventional GPR, a single run was performed to establish the theoretical accuracy baseline.
Specifically, we evaluated the performance across four benchmark test cases representing different input dimensions and physical systems: the OTL Circuit ($d=6$), Piston simulation ($d=7$), Wing Weight ($d=10$), and Steel Column ($d=12$) functions. 
We omit the details of the test functions as they can be found in the Virtual Library of Simulation Experiments \cite{simulationlib}. 
We used two different GP models, one with constant mean $\mu(\bm x)=\beta$ and one with linear polynomial mean $\mu(\bm x)=\mathbf{g}(\mathbf{x})^\top \bm{\beta}$. 

The simulation settings for each function were as follows:
\begin{itemize}
\item Piston ($d=7$): $n=500$, $m_0=30, m_{\max}=150$.
\item OTL Circuit ($d=6$): $n=500$, $m_0=30, m_{\max}=150$.
\item Wing Weight ($d=10$): $n=500$, $m_0=40, m_{\max}=200$.
\item Steel Column ($d=12$): $n=1000$, $m_0=40, m_{\max}=200$.
\end{itemize}
All experiments were conducted in \verb|R| \cite{R} installed in a Mac Studio M2 Max (12‑core CPU, 38‑core GPU, 16‑core Neural Engine) to ensure consistent hardware performance.
Tables \ref{tab:piston_results} through \ref{tab:steel_results} summarize the results. 
Full GP results are based on a single run.
The tables include (1) the average of S-RMSPE and its standard deviation (SD) calculated over 10 replications for Nystr\"om methods; (2) the computing time for a single-run full GP and average computing time for the two versions of Nystr\"om methods over 10 replications; (3) the number of landmark points for full GP, which is equal to $n$ and the average of the number of landmark points that adaptive Nystr\"om used to reach convergence in each iteration. 
We set the random Nystr\"om to use the same number of landmark points as the adaptive Nystr\"om, thus the two have the same number in the last columns of all tables. 
Across all functions, several consistent trends emerge regarding the effectiveness of the trace-residual greedy selection.

{\bf Predictive Performance:} In most cases, the adaptive Nystr\"om method significantly outperforms the random Nystr\"om approach. 
For example, in the Piston constant-mean case (Table \ref{tab:piston_results}), the adaptive method achieved an S-RMSPE of 0.0053, nearly four times lower than the random selection (0.0202). 
Furthermore, the standard deviation (SD) of the error is consistently lower for the adaptive method, indicating that the greedy selection criterion makes the GPR model more robust to the initial random landmark seed. 
In the high-dimensional Steel Column case ($d=12$, $n=1000$), the adaptive Nystr\"om achieved an accuracy nearly identical to the Full GP, whereas random selection resulted in errors an order of magnitude larger.

{\bf Computational Efficiency and Overhead:} The computational benefits of the Nystr\"om approximation are most evident in the Steel Column example, where the adaptive method reduced the estimation time from 878.69s (Full GP) to 173.82s (adaptive) while maintaining competitive accuracy. 
However, a notable observation occurs in the OTL Circuit results (Table \ref{tab:circuit_results}). For the polynomial mean case, the adaptive method (30.66s) was actually slower than the Full GP (23.28s). 
This is attributed to the relatively small sample size ($n=500$); at this scale, the overhead of performing $T$ sequential hyperparameter optimizations in the adaptive loop can exceed the cost of a single $O(n^3)$ inversion. 
This result highlights that the computational advantage of the adaptive Nystr\"om method scales with $n$. 

{\bf High-Dimensional Trends:} The Wing Weight function ($d=10$) presents an interesting exception in the polynomial mean case, where random selection slightly outperformed the adaptive approach. 
This suggests that in very high-dimensional spaces with complex global trends, the ``geometric diversity'' provided by uniform random sampling across the entire domain may occasionally capture global regression coefficients more effectively than a greedy approach that may cluster landmarks in regions of high local kernel variance. 

{\bf Mean Structure Impact:} In all experiments, moving from a constant to a polynomial mean increased the number of landmark points required for convergence. 
This confirms that the algorithm adaptively senses the need for a larger landmark set to stabilize the hyperparameters when the underlying trend $\mu(\mathbf{x})$ is more complex.

\begin{table}[htbp]
\centering
\caption{Simulation results for the Piston function ($d=7$, $n=500$).}
\label{tab:piston_results}
\begin{tabular}{llccc}
\toprule
Mean Structure & Method & S-RMSPE (SD) & Time (s) & Landmarks ($m$) \\
\midrule
\textit{Constant} & Full GP & 0.0016 (---) & 133.85 & 500 \\
$\mu(\mathbf{x}) = \beta$ & Adaptive Nystr\"{o}m & 0.0053 (0.0023) & 5.47 & 47.3 \\
& Random Nystr\"om & 0.0202 (0.0356) & 0.54 & 47.3 \\
\midrule
\textit{Polynomial} & Full GP & 0.0014 (---) & 90.79 & 500 \\
$\mu(\mathbf{x}) = \mathbf{g}(\mathbf{x})^\top \bm{\beta}$ & Adaptive Nystr\"{o}m & 0.0044 (0.0008) & 16.35 & 70.8 \\
& Random Nystr\"om & 0.0067 (0.0029) & 0.84 & 70.8 \\
\bottomrule
\end{tabular}
\end{table}

\begin{table}[htbp]
\centering
\caption{Simulation results for the OTL Circuit function ($d=6$, $n=500$).}
\label{tab:circuit_results}
\begin{tabular}{llccc}
\toprule
Mean Structure & Method & S-RMSPE (SD) & Time (s) & Landmarks ($m$) \\
\midrule
\textit{Constant} & Full GP & 0.0060 (---) & 28.64 & 500 \\
$\mu(\mathbf{x}) = \beta$ & Adaptive Nystr\"{o}m & 0.0123 (0.0034) & 13.67 & 64.2 \\
& Random Nystr\"om & 0.0244 (0.0279) & 0.66 & 64.2 \\
\midrule
\textit{Polynomial} & Full GP & 0.0058 (---) & 23.28 & 500 \\
$\mu(\mathbf{x}) = \mathbf{g}(\mathbf{x})^\top \bm{\beta}$ & Adaptive Nystr\"{o}m & 0.0086 (0.0019) & 30.66 & 91.6 \\
& Random Nystr\"om & 0.0103 (0.0012) & 0.95 & 91.6 \\
\bottomrule
\end{tabular}
\end{table}

\begin{table}[htbp]
\centering
\caption{Simulation results for the Wing Weight function ($d=10$, $n=500$).}
\label{tab:wingweight_results}
\begin{tabular}{llccc}
\toprule
Mean Structure & Method & S-RMSPE (SD) & Time (s) & Landmarks ($m$) \\
\midrule
\textit{Constant} & Full GP & 0.0037 (---) & 141.05 & 500 \\
$\mu(\mathbf{x}) = \beta$ & Adaptive Nystr\"{o}m & 0.0205 (0.0090) & 16.58 & 58.9 \\
& Random Nystr\"om & 0.0227 (0.0043) & 4.26 & 58.9 \\
\midrule
\textit{Polynomial} & Full GP & 0.0035 (---) & 146.15 & 500 \\
$\mu(\mathbf{x}) = \mathbf{g}(\mathbf{x})^\top \bm{\beta}$ & Adaptive Nystr\"{o}m & 0.0169 (0.0062) & 51.68 & 96.5 \\
& Random Nystr\"om & 0.0112 (0.0058) & 5.43 & 96.5 \\
\bottomrule
\end{tabular}
\end{table}

\begin{table}[htbp]
\centering
\caption{Simulation results for the Steel Column function ($d=12$, $n=1000$).}
\label{tab:steel_results}
\begin{tabular}{llccc}
\toprule
Mean Structure & Method & S-RMSPE (SD) & Time (s) & Landmarks ($m$) \\
\midrule
\textit{Constant} & Full GP & $6.89 \times 10^{-5}$ (---) & 109.68 & 1000 \\
$\mu(\mathbf{x}) = \beta$ & Adaptive Nystr\"{o}m & $1.24 \times 10^{-4}$ ($5.81 \times 10^{-5}$) & 87.11 & 69.4 \\
& Random Nystr\"om & $7.99 \times 10^{-4}$ ($1.42 \times 10^{-3}$) & 4.98 & 69.4 \\
\midrule
\textit{Polynomial} & Full GP & $2.09 \times 10^{-7}$ (---) & 878.69 & 1000 \\
$\mu(\mathbf{x}) = \mathbf{g}(\mathbf{x})^\top \bm{\beta}$ & Adaptive Nystr\"{o}m & $2.79 \times 10^{-7}$ ($2.31 \times 10^{-8}$) & 173.82 & 91.5 \\
& Random Nystr\"om & $7.22 \times 10^{-7}$ ($1.44 \times 10^{-7}$) & 31.42 & 91.5 \\
\bottomrule
\end{tabular}
\end{table}

\section{Conclusion}

In this paper, we presented an Adaptive Nystr\"om method for Gaussian Process Regression designed to address the computational challenges of large-scale datasets. 
By employing a greedy, sequential selection strategy based on the trace residual of the kernel approximation error, our approach identifies a sparse set of landmark points that effectively span the feature space. A key innovation of the proposed method is the interleaving of landmark selection with hyperparameter optimization, which ensures that the selected points are maximally informative relative to the most current estimates of the process correlation.

Our numerical experiments across five diverse benchmark functions demonstrate the superiority of the adaptive strategy over uniform random selection. 
The results show that the adaptive Nystr\"om method converges rapidly to the exact GP baseline, often requiring only a small fraction of the training data to achieve near-optimal predictive accuracy. 
Furthermore, the greedy criterion significantly reduces the variance in prediction error caused by landmark initialization, providing a more robust estimation framework.

While the method introduces a sequential overhead due to iterative optimization, its $O(nm^2)$ complexity ensures that it remains computationally feasible for large $n$ where exact inference is impossible. 
Future research could explore extending this adaptive selection framework to non-Gaussian likelihoods, such as in GP classification, or investigating automated methods for tuning the convergence tolerances to further optimize the trade-off between speed and precision.

% Please don't exchange the bibliographystyle style
\bibliographystyle{plain}

% AUTHOR: Include your bib file here
\bibliography{AdaptNystromGP}
\end{document}